\documentclass[preprint2]{emulateapj}

\newcommand{\etal}{{et\,al.}}

\def\unit #1{\,{\rm #1}} 
\def\kmps{\unit{km \, s^{-1}}}

\shorttitle{Pseudobulges in S0 Galaxies}
\shortauthors{Vaghmare et al.}

\begin{document}

\title{A Spitzer  Study of Pseudobulges in S0 Galaxies : Secular
Evolution of Disks}

\author{Kaustubh Vaghmare\altaffilmark{1}, 
Sudhanshu Barway\altaffilmark{2} and
Ajit Kembhavi\altaffilmark{1}
}

\altaffiltext{1}{IUCAA, Post Bag 4, Ganeshkhind, Pune 411007, India	;
kaustubh@iucaa.ernet.in, akk@iucaa.ernet.in} 
\altaffiltext{2}{South African Astronomical Observatory, P.O. Box 9, 7935,
Observatory, Cape Town, South Africa; barway@saao.ac.za }

\begin{abstract}
In this Letter, we present a systematic study of lenticular (S0) galaxies based
on mid-infrared imaging data on 185 objects taken using the {\it {\it Spitzer} }
Infra Red Array Camera. We identify the S0s hosting pseudobulges based on the
position of the bulge on the Kormendy diagram and the S\'{e}rsic index of the
bulge. We find that pseudobulges preferentially occur in the fainter luminosity
class (defined as having total $K$-band absolute magnitude $M_{\rm{K}}$
fainter than $-22.66$ in the AB system). We present relations between bulge and
disk parameters obtained as a function of the bulge type. The disks in the
pseudobulge hosting galaxies are found to have distinct trends on the 
$r_{\rm{e}}-r_{\rm{d}}$  and $\mu_{\rm{d}} (0) - r_{\rm{d}}$
correlations compared to those in galaxies with classical bulges. We show that
the disks of pseudobulge hosts possess on average a smaller scale length and
have a fainter central surface brightness than  their counterparts occurring in
classical bulge hosting galaxies. The differences found for discs in pseudobulge
and classical bulge hosting galaxies may be a consequence of the different
processes creating the central mass concentrations.

\end{abstract}
  

\section{Introduction}

Bulges are a central piece of the puzzle of galaxy formation and evolution.
Recent work - theoretical and observational - has revealed that bulges come in
two flavors: those thought to have formed through violent processes such as
hierarchical clustering or major mergers and referred to as ``classical bulges"
and those thought to have formed through secular processes and referred to as
``pseudobulges" (Kormendy \& Kennicutt 2004). \nocite{Kormendy2004} Many
studies have been carried out proposing and examining various criteria for
identifying these two classes and distinguishing between them. For example,
Carollo et al. (1998) \nocite{Carollo98} identify pseudobulges based
on the presence of nuclear structure as seen in images taken using the
{\it Hubble Space Telescope} (HST). Fisher \& Drory (2008)
\nocite{FisherDrory2008} employed a similar method and proposed a classification
of bulges into ``classical'' and  ``pseudo'' based on the S\'{e}rsic index of
the bulge. Carollo et al. (2001) \nocite{Carollo2001} find a difference in the
average optical-near-infrared $V-H$ color of the two kinds of bulges. The two
types exhibit different behavior on various well known correlations between the
structural parameters of the galaxy. The studies also show a smooth (rather than
sharp) transition from one type of bulge to the other, possibly implying the
existence of bulges with a mixture of properties (Gadotti 2009; Fisher \& Drory
2010).\nocite{Gadotti2009} \nocite{FisherDrory2010}

The above studies have focussed on bulge dichotomy using samples comprising of
all galaxy types with a bulge and a disk but it can also be interesting to study
this dichotomy within the same morphological class. In the present work, we
consider S0 galaxies which form an intermediate transition class between
ellipticals and spiral galaxies on the Hubble tuning fork diagram (Hubble 1936).
\nocite{Hubble1936} Recent studies hint at existence of sub-populations within
S0 galaxies, with different properties and formation histories. Barway \etal \ 
(2007, 2009, 2011) \nocite{BWK2009} \nocite{BWK2007} \nocite{BWK2011}have
proposed a division of S0 galaxies into bright and faint classes using
a dividing $K$-band absolute magnitude of $-24.5$ in the Vega
system. The properties and behavior with respect to various well known
correlations between morphological parameters are different for the two
classes. For example, Barway et al. (2007) find a strong
positive correlation between the bulge effective radius $r_{\rm{e}}$ and
the disk scale length $r_{\rm{d}}$ for fainter 
galaxies while a weak anti-correlation is found in the case of bright galaxies. 
Barway et al. (2009) report distinct trends for bright and faint S0s on
correlations such as the Kormendy relation, the photometric plane etc. The
differences can be interpreted as the dependence of the formation history of the
galaxies on the luminosity and the environment of the galaxies: the more
luminous S0 galaxies should have formed through more violent processes while the
faint ones through secular evolution.

If pseudobulges in S0 galaxies have formed through processes wherein
the disk material rearranges itself to form the bulge, it is expected that the
disk itself undergoes a change in its properties when giving rise to
the bulge. In this Letter, we have carried out a mid-infrared study of a large
sample of S0 galaxies to examine the properties of the bulges
and the host disks. The Letter is organized as
follows. Section 2 describes the sample selection and the data. Section 3
describes the various results found and Section 4 summarizes the findings and
their implications. Throughout this Letter, we have assumed the standard
concordance cosmology i.e. $H_{\rm 0} = 70\,\kmps\,\unit{Mpc^{-1}}$,
$\Omega_{\rm{m}} = 0.3$ and $\Omega_{\Lambda} = 0.7$.

\section{Sample and Data Analysis}

To assemble a statistically significant sample of S0 galaxies, we
started with the RC3 catalog (de Vaucouleurs et al. 1991) \nocite{deVauc91} and
identified all galaxies classified as S0 (with numerical Hubble stage
$-3 \le T \le 0$). The RC3 contains 3657 such galaxies. We impose a magnitude
cut with total apparent $B$ magnitude $B_T < 14.0$ which reduces the sample
size to 1031 galaxies. For these galaxies, we obtained the total apparent
$K_{\rm{s}}$ magnitudes from the Two Micron All Sky Survey (2MASS) All Extended
Source Catalog and converted these to absolute magnitudes using redshifts
obtained from the NASA Extragalactic Database
(NED).\footnote{http://ned.ipac.caltech.edu/}

The mass content of a galaxy is dominated by low mass stars and radiation at
infrared wavelengths takes into account their contribution and is also
relatively free from effects of dust extinction. It is therefore appropriate
to use data at near- or mid-infrared wavelengths for the study of galaxy
morphology. We used the imaging data at 3.6 $\mu$m taken using the Infrared
Array Camera (IRAC) on board the {\it Spitzer  Space Telescope}. IRAC
offers deep images at mid-infrared wavelengths free from the problems of sky
emission that plague ground based observations. 

We crossmatched the above magnitude limited sample with the data available
in the Spitzer Heritage Archive (SHA)\footnote{The {\it Spitzer} 
Heritage Archive is maintained by the Spitzer Science Centre and is a
public interface to all archival data taken using the three instruments on board
the {\it Spitzer Space Telescope.} \linebreak (
http://irsa.ipac.caltech.edu/applications/Spitzer/SHA/)} and
found that imaging data for 247 galaxies were available. {\it Spitzer}  takes
dithered observations of the extended objects, which were downloaded from the
SHA and coadded to a final mosaic using the MOsaicking and Point EXtraction
(MOPEX) tool provided by the {\it Spitzer}  Science Centre. MOPEX performs the
necessary preprocessing steps which include accounting for optical distortions,
image projection and outlier pixel rejection to create a final science mosaic. 

In order to derive the structural parameters of various galaxy components, we
employed the technique of two-dimensional decomposition of galaxy light. We used
the program GALFIT (Peng \etal \  2002) \nocite{Peng2002} for this purpose. In
the first run, we fitted all galaxies with a bulge and a disk component using a
S\'{e}rsic (S\'{e}rsic 1968) \nocite{Sersic68} and an exponential profile
respectively. For galaxies where the residual image  obtained by subtracting the
point spread function convolved best-fit model from the observed image revealed
a bar, a second run of fitting was performed by adding another S\'{e}rsic
component to describe the bar. The bulge properties thus determined are free
from the systematics that arise if a bar which is present remains unaccounted
(Gadotti 2008; Laurikainen et al. 2005). \nocite{Gadotti2008}
\nocite{Laurikainen2005} During the course of the decomposition,
galaxy images with poor signal-to-noise ratio and/or complicated morphological
features such as tidal tails due to recent mergers were discarded from the
sample. The final sample comprises 185 galaxies with a median redshift of
$\sim0.005$ and a standard deviation of $\sim0.002$. Further, the sample size is
limited only by the availability of good quality data for non-interacting /
undisturbed S0s in the SHA. The S0s were classified as bright and faint
following the criterion by Barway \etal \  (2007, 2009, 2011), who proposed a
dividing Vega magnitude of $-24.5$ on the 2MASS $K_{\rm{s}}$ absolute magnitude
scale for this purpose. We transformed this division line to the AB system using
$K_{\rm{s}} (\rm{AB}) = K_{\rm{s}} (\rm{Vega}) + 1.84$ as suggested by
Mu{\~n}oz-Mateos \etal \  (2009). The sample comprises 37 bright and 148
faint galaxies as determined using the new dividing magnitude of $-22.66$.
\nocite{Munoz2009}

We have used the AB magnitude system (Oke \& Gunn 1983)
\nocite{OkeGunn83} throughout this Letter and where necessary,
magnitudes obtained from various catalogs have been transformed to this
system.

\section{Pseudobulges in S0 Galaxies}

\subsection{Identifying Pseudobulges} 
Fisher \& Drory (2008) \nocite{FisherDrory2008} classified as pseudobulges
those bulges which had the presence of nuclear spiral arms, bars and/or rings as
seen in the high resolution near \linebreak $V$ - band images taken using the
{\it Hubble Space Telescope} (see also Carollo et al. 1998). The rest were
classified as classical bulges. They found that classical bulges have S\'{e}rsic
index $n > 2$ and follow the same relations between various structural
parameters as elliptical galaxies. But these relations were not obeyed by the
pseudobulges which are found to have $n < 2$. 

It appears from above that the S\'{e}rsic index can be used to classify the
bulges into the two types, classifying the bulges with $n<2$ as pseudobulges and
those with $n>2$ as classical. If we use this criterion to classify the bulges
of the S0 galaxies in our sample into these two types, we have 111 classical
bulges and 74 pseudobulges. Only a small fraction (13/74) of pseudobulges belong
to the bright S0s with the rest (61/74) being associated with faint galaxies.
However, the typical error bars on $n$ can be quite significant. Our experience
with various data shows that this error can be as large as 20\% while Gadotti
(2008) \nocite{Gadotti2008} reports errors as large as 0.5. The parameters of
the S\'{e}rsic profile $n$ and $r_{\rm{e}}$ are coupled which leads to an
additional uncertainty in $n$ (see Trujillo et al. 2001 and references therein)
\nocite{Trujillo2001}. Thus classifying bulges using $n$ {\it alone} can lead to
ambiguity.

Following Fisher \& Drory (2008), we tried to use high resolution {\it HST}
images for constraining the classification of the bulges by searching for
evidence of nuclear structure. For our S0 sample, high resolution {\it HST}
imaging data were available for only 110 of the 185 galaxies in various optical
bands. In the absence of a homogeneous sample of {\it HST} images, it was not
possible to impose constraints using this method. 

\begin{figure}
\plotone{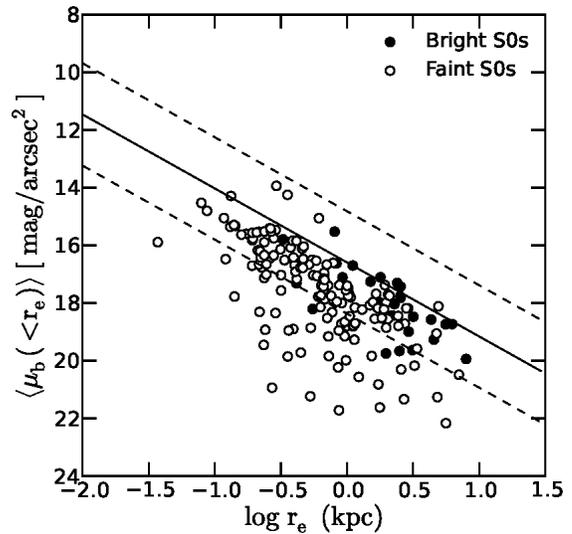}
\caption{Kormendy relation with filled and open circles representing bright and
faint S0s respectively. The solid line is the best-fit line to Coma cluster
ellipticals while the dashed lines mark the 3-$\sigma$ limits.}
\label{KormendyBF}
\end{figure}

A different approach used by Gadotti (2009), \nocite{Gadotti2009} which avoids
the difficulties associated with using $n$ alone, involves classifying bulges
based on their position on the Kormendy diagram (Kormendy 1977). This is a plot
of the average surface brightness of the bulge within its effective radius
$\left<\rm{\mu_b} (<r_{\rm{e}})\right>$ against the logarithm of the effective
radius $r_{\rm{e}}$. The elliptical galaxies are known to obey a tight,
linear correlation on this diagram. Classical bulges resemble elliptical
galaxies and obey a similar relation while pseudobulges lie away from it. Any
bulge which deviates by more than three times the rms (root mean square) scatter
from the best-fit relation for ellipticals is classified as a pseudobulge by
Gadotti (2009).

The Kormendy plot for our sample is shown in Figure
\ref{KormendyBF} with filled and empty circles respectively denoting bright and
faint galaxies. Since the classification of a bulge depends on its deviation
from a best-fit line to the ellipticals, the decomposition data from a study by
Khosroshahi et al. (2000), \nocite{Khosroshahi2000} of Coma cluster ellipticals
done in the $K$-band, was used for comparison with our data. The magnitudes were
transformed to the 3.6 $\mu$m band using the relation $K - m_{3.6} =
0.1$ (Toloba et al. 2012; Falc{\'o}n-Barroso et al. 2011) \nocite{Toloba2012}
\nocite{Falcon2011} and to the AB magnitude system by adding 1.84 following
Mu{\~n}oz-Mateos et al. (2009). The best-fit line for these ellipticals is

\[ 
\left<\rm{\mu}_{\rm{b}} (<r_{\rm{e}})\right> = (2.567 \pm 0.511)\ \log\
r_{\rm{e}} + 16.595 \pm 0.296 \rm{.} 
\]

\noindent The rms scatter in $\left<\rm{\mu_b} (<r_{\rm{e}})\right>$ for the
Coma cluster ellipticals is 0.592. The best-fit line is shown in Figure
\ref{KormendyBF}. The dashed lines in the diagram enclose a region where all
points within three times the rms scatter can be found. Following Gadotti
(2009), points below this region can be classified as pseudobulges and there are
50 such points. 

Interestingly, most pseudobulge hosting S0s thus classified are faint. Only
three of the 37 bright galaxies are classified as pseudobulge hosts based on
their position on the Kormendy diagram. If we assume that both bright and faint
S0s have the same fraction (32\%) of pseudobulges, we can estimate the
probability of finding only three pseudobulges among bright S0s using the
binomial distribution. This probability is found to be $\sim 10^{-4}$,
suggesting that it is highly unlikely that bright and faint galaxies have the
same chance of hosting a pseudobulge. 

A set of prescriptions has been suggested by Kormendy \& Kennicutt (2004) to
put constraints on the bulge classification. The more constraints a bulge
satisfies, the more secure its classification becomes. Apart from examining
{\it HST} images for various nuclear structures, one could investigate the
$D_{\rm{n}} (4000)$ indices (Gadotti 2009) to check for recent star
formation as pseudobulges are known to contain a recently formed stellar
population and may even be forming stars actively. However, $D_{\rm{n}}
(4000)$ measurements, which are obtainable from the Sloan Digital Sky Survey
(SDSS) data products, are again not available for a large fraction of the S0
galaxies in our sample. The SDSS is constrained to a specific portion of the
sky while our sample is an all-sky sample.

In the absence of the above indicators to have more secure classification of
bulges in our S0 sample, we chose to impose a cut off on the S\'{e}rsic index
$n$  for the pseudobulge candidates obtained from the Kormendy diagram as
explained above. We accept as a pseudobulge only those bulges which lie below
the $3 \sigma$ line and have $n<2$.  This reduces the final sample of
pseudobulges to 27 (of which two are bright). All other bulges are classified as
``classical''. Where available, we have used {\it HST} images to confirm the
presence of nuclear structure in these galaxies. Further, wherever possible, we
have checked {\it HST} images of those galaxies classified as classical bulge
hosts and verified that these galaxies do not show any nuclear features. We
believe that our pseudobulge  identification  is reasonably secure. The
subsequent sections are based on this sample of 27 pseudobulges and 158
classical bulges.

\subsection{Disk Correlations as Function of Bulge Type}

\begin{figure}
\plotone{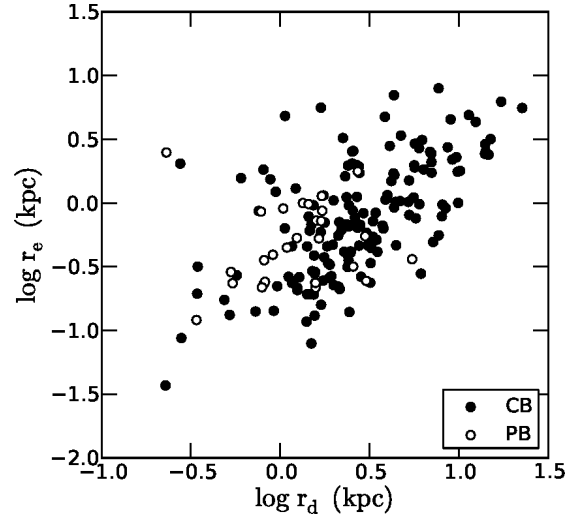}
\caption{Plot of $r_{\rm{e}}$ vs $r_{\rm{d}}$ with filled and empty circles
denoting classical bulges (CB) and pseudobulges (PB), respectively.}
\label{rerd}
\end{figure}

\begin{figure}
\plotone{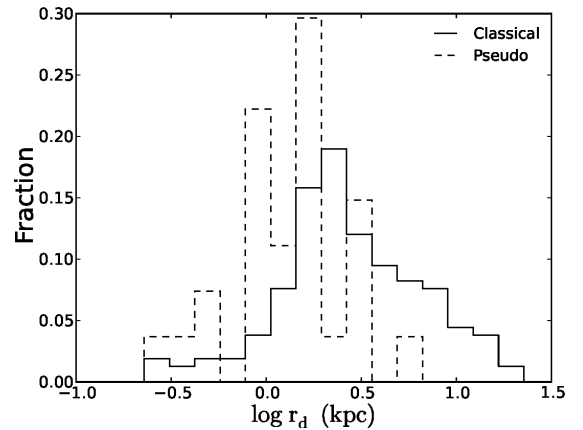}
\caption{Scaled histogram showing the distribution of $\log
r_{\rm{d}}$. The solid and dotted lines represent distributions for
classical and pseudobulges, respectively.}
\label{Hist_rd}
\end{figure}

If pseudobulges are formed by processes within the disk, one expects a
correlation between the bulge and disk scale lengths (Courteau et al. 1996),
\nocite{Courteau96} which is predicted by secular bulge formation models
(Martinet 1995; Combes 2000; see Carollo et al. 1999 for comprehensive review
articles). \nocite{Martinet95} \nocite{Combes2000} \nocite{Carollo99} Also, one
expects an imprint of the pseudobulge formation processes on correlations
involving disk parameters.

\begin{figure}
\plotone{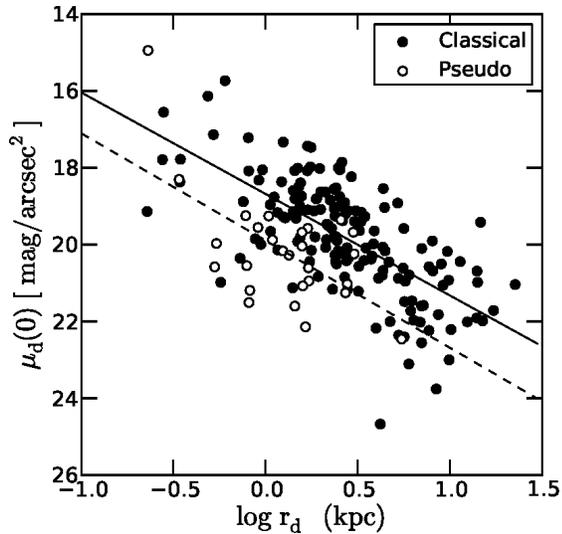}
\caption{Plot of disk central surface brightness as a function its scale
length. The filled and empty circles represent the classical and pseudobulges,
respectively. The best-fit straight lines to them are the solid and the
dashed lines.}
\label{Disk_mu0_rd}
\end{figure}

In Figure \ref{rerd} we plot bulge effective radius $r_{\rm{e}}$ as a function
of the disk scale length $r_{\rm{d}}$. The filled and empty circles respectively
denote classical bulge hosting S0s and pseudobulge hosts.
The empty circles denoting the pseudobulge hosts are
shifted toward the left relative to the filled circles which represent
classical bulges. This indicates that the disk scale length is smaller on
average for pseudobulge hosts than classical bulge hosts. This can be seen in
Figure \ref{Hist_rd}. The mean disk scale length for pseudobulge hosts is
$1.6\ \rm{kpc}$ while that of classical bulges is $4\ \rm{kpc}$. A $t$-test
rules out at greater than 99.9\% confidence, the possibility of this difference
arising from random chance alone. This indicates that pseudobulges
preferentially occur in smaller disks. The nature of the bulge depends on the
formation history, in particular on whether the bulge is a product of a process
such as a major merger or whether it has formed from secular evolution. If we
assume that there is no reason to believe that only galaxies with larger disks
should undergo such a merger it might be possible that the lower scale length of
the disk in the case of a pseudobulge is due to the processes within it that
grew the bulge. 

If secular processes that grow pseudobulges leave behind an imprint on the
progenitor disks, one should be able to find them on correlations
involving disk parameters. In Figure \ref{Disk_mu0_rd}, we plot the central
surface brightness of the disk $\mu_{\rm{d}}\ (0)$ as a function of
$r_{\rm{d}}$. A correlation overall is seen, with fainter central
brightness corresponding to a larger disk scale length, but there is a clear
offset between disks of pseudobulge hosts and those of classical bulge hosts.
For a given $r_{\rm{d}}$, the disk hosting a pseudobulge is fainter
than the one hosting a classical bulge. 

We have performed two-sample Kolmogorov - Smirnov tests to compare the
distributions of  central surface brightness, absolute magnitude of disk, and
$r_{\rm{d}}$ for classical and pseudobulge hosting S0s. We find that these
samples of classical bulge hosts and pseudobulge hosts could not have been drawn
from the same parent population, with at least 99.9\% confidence. 

In his paper, Gadotti (2009) finds that the disks of pseudobulge
hosts are more extended and have a fainter surface brightness compared to
those of classical bulge hosts but the overlap of the two kinds of galaxies is
significant. It is important to note that Gadotti (2009) focusses on bulge
dichotomy for a sample comprising of a mixture of different morphological types
while our sample comprises S0s alone.

\section{Discussion}

We have presented the first systematic study of pseudobulges in S0 galaxies with
emphasis on signatures of their evolutionary processes on their progenitor
disks. We use the position of the bulge on the Kormendy diagram as an initial
classification criterion for determining the nature of the bulge. To make our
classifications more secure we have {\it also} used the criterion proposed by
Fisher \& Drory (2008) which involves using a division line of $n=2$ on the
S\'{e}rsic index scale. There are 27 pseudobulges in our sample of which two
belong to bright S0s. Thus pseudobulges occur preferentially in fainter galaxies
($M_{\rm{K}} > -22.66$, AB system). Using plots between bulge effective
radius versus disk scale length and disk central surface brightness versus disk
scale length, we demonstrate that distinct trends are followed by classical and
pseudobulge host S0 galaxies.

The disks of pseudobulge hosting S0s have on average smaller scale lengths,
lower central surface brightness and luminosity. One can either interpret this
as pseudobulges preferentially occurring in a different population of disks or
that these signatures are an imprint of the processes that drove the pseudobulge
growth. It is not clear how to explain this finding using one of the known
mechanisms for pseudobulge formation. For example, transforming a spiral with a
pseudobulge via ram pressure stripping (Abadi et al. 1999) \nocite{Abadi99} can
perhaps cause disk fading but may not cause a change in the scale length
resulting in the lower $r_{\rm{d}}$ reported here. Perhaps a combination of
processes may be able to explain the findings reported in this Letter.

In a future work, we will present a study of the current sample in greater
detail and search for further signatures of the underlying processes involved in
bulge and pseudobulge formation in S0 galaxies.

\acknowledgements 
We thank the anonymous referee for insightful comments that have improved both
the content and the presentation of this Letter.

We express our sincere gratitude to Yogesh Wadadekar for his
comments and suggestions. Various statistical tests used in the Letter have been
carried out using 
AstroStat\footnote{http://vo.iucaa.ernet.in/$\sim$voi/AstroStat.html},
a statistical package developed by Virtual Observatory - India. We also
acknowledge the use of IPython (P\'{e}rez \& Granger, 2007). \nocite{Perez2007}

This work is based on observations made with the
{\it Spitzer Space Telescope}, which is operated by the Jet Propulsion
Laboratory,
California Institute of Technology under a contract with NASA. We acknowledge
the use of the HyperLeda database (http://leda.univ-lyon1.fr). 

This material is based upon work supported financially by the National Research
Foundation (NRF). Any opinions, findings and conclusions or recommendations
expressed in this material are those of the author(s) and therefore the NRF does
not accept any liability in regard thereto.

\bibliographystyle{apj}

\end{document}